\documentclass[twocolumn, secnumarabic,amssymb, nobibnotes, aps, prd, superscriptaddress]{revtex4-1} 
\usepackage{amsmath}
\usepackage[breaklinks=true]{hyperref}
\usepackage{cleveref}

\usepackage{graphicx}
\usepackage[font=normal]{subfig}

\usepackage{booktabs}
\usepackage{tabularx}

\usepackage{enumitem}
\setlist[itemize]{align=parleft,left=0pt..1em}

\begin{document}

\title{Markov Chain Monte-Carlo Enhanced Variational Quantum Algorithms}

\author{Taylor L. Patti}
\email[]{taylorpatti@g.harvard.edu}
\affiliation{Department of Physics, Harvard University, Cambridge, Massachusetts 02138, USA}
\affiliation{IBM Quantum, IBM T.J. Watson Research Center, Yorktown Heights, NY 10598, USA}

\author{Omar Shehab}
\affiliation{IBM Quantum, IBM T.J. Watson Research Center, Yorktown Heights, NY 10598, USA}

\author{Khadijeh Najafi}
\affiliation{Department of Physics, Harvard University, Cambridge, Massachusetts 02138, USA}
\affiliation{IBM Quantum, IBM T.J. Watson Research Center, Yorktown Heights, NY 10598, USA}

\author{Susanne F. Yelin}
\affiliation{Department of Physics, Harvard University, Cambridge, Massachusetts 02138, USA}

\begin{abstract}
Variational quantum algorithms are poised to have significant impact on high-dimensional optimization, with applications in classical combinatorics, quantum chemistry, and condensed matter. Nevertheless, the optimization landscape of these algorithms is generally nonconvex, causing suboptimal solutions due to convergence to local, rather than global, minima. In this work, we introduce a variational quantum algorithm that uses classical Markov chain Monte Carlo techniques to provably converge to global minima. These performance guarantees are derived from the ergodicity of our algorithm's state space and enable us to place analytic bounds on its time-complexity. We demonstrate both the effectiveness of our technique and the validity of our analysis through quantum circuit simulations for MaxCut instances, solving these problems deterministically and with perfect accuracy. Our technique stands to broadly enrich the field of variational quantum algorithms, improving and guaranteeing the performance of these promising, yet often heuristic, methods.
\end{abstract}

\maketitle

\section{Introduction}
\label{sec:intro}

Since\footnote{While finalizing this manuscript, we became aware of another work applying Markov Chain Monte-Carlo technique in quantum algorithms \cite{Mazzola-mcmc-2021}. However, we differentiate our work by targeting near-term quantum algorithms and providing the proof of ergodicity.} the advent of the Variational Quantum Eigensolver (VQE) \cite{mcclean2016theory, peruzzo2014variational} and Quantum Approximate Optimization Algorithm (QAOA) \cite{farhi2014quantum}, quantum algorithms that function in tandem with classical machine learning have garnered great interest. These variational quantum algorithms (VQAs) typically harness some form of classical gradient descent to tackle a large-scale optimization problem on the exponential state space of quantum hardware \cite{Cerezo2021, lavrijsen2020classical, cerezo2021variational}. Applications of these methods have included the optimization of NP-hard combinatorial problems \cite{garey2002computers, Nannicini2019, Braine2021, Patti2021, Fuller2021}, the identification of eigenstates and energies in quantum chemistry applications \cite{mcardle2018quantum, kandala2017hardware, Grimsley2019}, and the study of condensed matter systems \cite{ritter2019near, vogt2020preparing, Zhang2021}. Much like their classical counterparts, the above near-term quantum algorithms can be plagued by nonconvex optimization landscapes, causing them to converge to suboptimal minima \cite{Lee2021}. A variety of techniques have been suggested to address this issue in NP-hard combinatorial optimization problems, such as: ``warm starting'' proceedures \cite{beaulieu2021max, egger2021warm, van2021}, composition with classical neural networks \cite{Rivera-Dean2021}, multibasis encodings with bistable convergence \cite{Patti2021}, and other  techniques \cite{Fuller2021, harwood2021improving, shehab2019noise}. However, these methods offer few provable optimization guarantees of practical utility. While optimization landscapes are known to become more convex with high-depth \cite{Lee2021}, the adverse effect of quantum noise \cite{bravyi2018quantum} and barren plateaus \cite{mcclean2018barren, Patti2020,Marrero2020,holmes2021connecting,Cerezo2020} on deep quantum networks is well-documented.

In order to avoid the local minima convergence that plagues VQAs, we introduce MCMC-VQA, a technique that adapts the ergodic exploration of classical Markov chain Monte Carlo (MCMC) to guarantee the global convergence of quantum algorithms. As samples of ergodic systems are representative of their underlying probability distribution, an ergodic VQA necessarily yields a sample that contains states near the global minimum. In this work, we focus on the Metropolis-Hastings algorithm due to its success in high-dimensional spaces and suitability for unnormalized probability distributions \cite{Metropolis1953}. MCMC-VQA utilizes modified VQAs and their statistics as the Metropolis-Hastings transition kernels and quantum state energies as state likelihoods. These quantities are then used to determine the viability of parameter updates. Our algorithm requires no increase in quantum overhead and only a minimal increase classical overhead. MCMC-VQA represents a time-discrete, space-continuous Markov chain, as the algorithm progresses in discrete VQA epochs while training a continuous-parameter quantum circuit. It can also be classified as a form of Stochastic Gradient Descent MCMC \cite{Robbins1951, Nemeth2021}. Although in this work we focus on VQE \cite{peruzzo2014variational}, our techniques are readily applicable to a wide array of quantum machine learning applications.

While other works have introduced quantum subroutines for classical MCMC methods that offer a quadratic speedup for random walks \cite{Szegedy2004,Temme2011, Lemieux2020} and sampling \cite{Montanaro2015, Cornelissen2021}, this manuscript takes the opposite approach by designing a classical MCMC subroutine for quantum algorithms. Likewise, while classical MCMC methods have been used to \textit{simulate} quantum computing routines \cite{Wang2016, Medvidovic2021}, our work uses classical MCMC to \textit{enhance} quantum algorithms on quantum hardware.

\begin{figure*}
\includegraphics[width=0.9\textwidth]{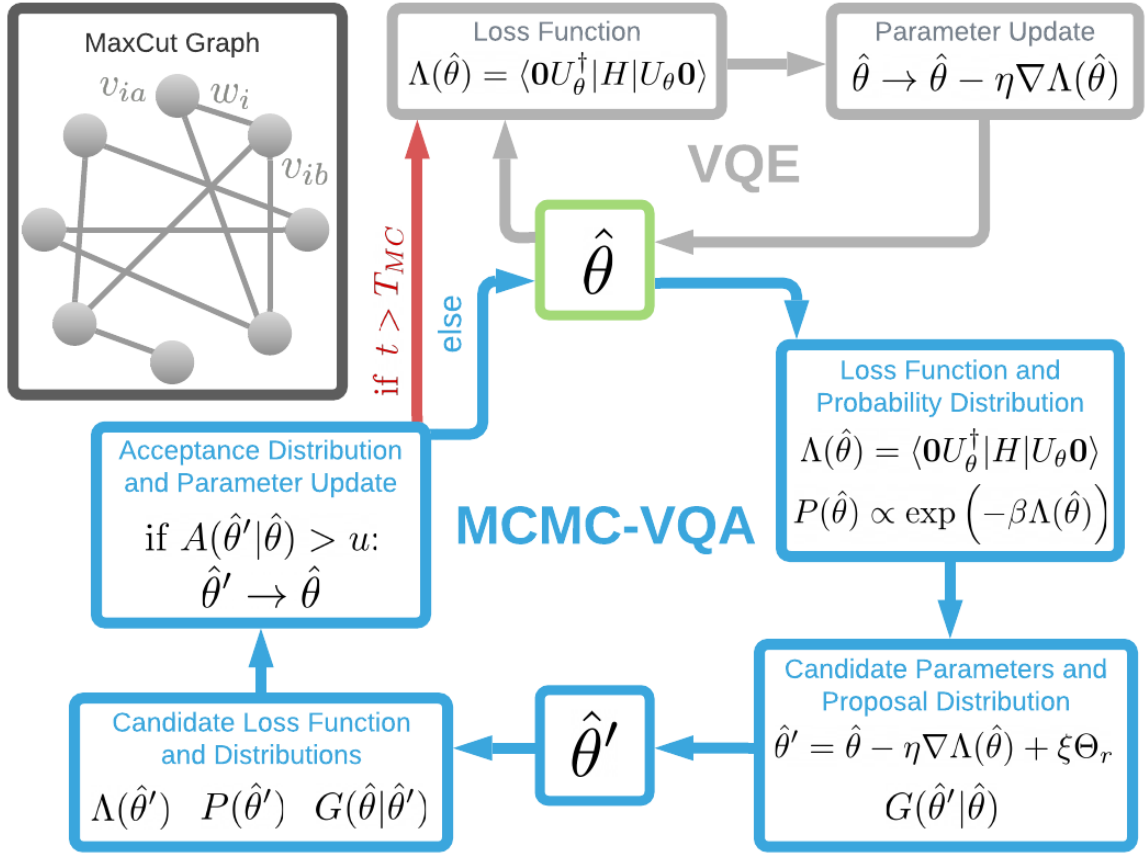}
\caption{Diagram of a random graph for MaxCut, VQE, and MCMC-VQA. \textbf{Random graphs} (black, Secs.\ \ref{sec:intro} and \ref{sec:methods}) in this work are generated with normally distributed edge weights $w_i$. The objective is to minimize Eq.\ \ref{eq:maxcut} by optimally assigning each pair of vertices $v_{ia}$, $v_{ib} \in \{-1,1\}$. MaxCut can be solved on a quantum computer by mapping $v_{ia}$, $v_{ib} \rightarrow \sigma_{ia}$, $\sigma_{ib}$ and minimizing the corresponding $H$. See Sec.\ \ref{sec:methods} for graph details. \textbf{VQE} (gray, Sec.\ \ref{sec:intro}) minimizes the loss function for each $\hat{\theta}$ by calculating the expectation value $\Lambda(\hat{\theta})$ and updating $\hat{\theta}$ with gradient descent using $\nabla \Lambda(\hat{\theta})$. \textbf{MCMC-VQA} (blue, Sec.\ \ref{sec:results}) uses gradient descent with $\nabla \Lambda(\hat{\theta})$ and random noise $\xi \Theta_r$ to produce candidate state $\hat{\theta}'$, but also calculates probability distributions $P(\hat{\theta})$ and $P(\hat{\theta}')$, as well as proposal distributions $G(\hat{\theta}'|\hat{\theta})$ and $G(\hat{\theta}|\hat{\theta}')$. Using these distributions, the acceptance distribution $A(\hat{\theta}'|\hat{\theta})$ is calculated and compared to random uniform sample $u \sim U(0,1)$. If $A(\hat{\theta}'|\hat{\theta}) > u$, then $\hat{\theta}' \rightarrow \hat{\theta}$. Otherwise, the MCMC-VQA algorithm restarts with the original $\hat{\theta}$. (Red) after the maximum number of MCMC-VQA epochs $T_\text{MC}$ have occurred, the sampled parameters with the lowest loss, $\hat{\theta}_\text{min}$, are selected and the optimization completes with a closing sequence of VQE epochs.}
\label{fig:1}
\end{figure*}

 We briefly review VQAs, focusing on VQE (Fig.\ \ref{fig:1}, gray) for quantum optimization for MaxCut problems. This choice of application is motivated by the ample nonconvexity of the corresponding quadratic loss functions \cite{Patti2021, Lee2021}. VQAs are parameterized by input states $|\psi \rangle$ and quantum circuit unitaries $U_t = U(\hat{\theta}_t)$, where $\hat{\theta}_t$ are the variable parameters learned during epoch $t-1$. Without loss of generality, we choose the $n$-qubit input state as $|\mathbf{0} \rangle = \prod_{i=0}^{n-1} |0\rangle$ such that the output state is entirely defined by $\hat{\theta}$ and assume that the initial parameters $\hat{\theta}_0$ are randomly selected at the start of each new sequence of epochs.
 
MaxCut is a partitioning problem on undirected graphs $G$ (Fig.\ \ref{fig:1}, black), where edges $\omega_i$ connect pairs of vertices $v_{ia}$, $v_{ib}$ \cite{Commander2009}. The goal is to optimally assign all vertices $v_{ia}$, $v_{ib} \in \{-1,1\}$, so as to maximize the objective function
 
 \begin{equation}
\textrm{maximize} \hspace{0.4cm} \frac{1}{2} \sum_{i} w_{i} \left(1-v_{ia} v_{ib} \right).
\label{eq:maxcut}
\end{equation}

\noindent In this work, we will consider a generalized form of the problem known as \textit{weighted} MaxCut, in which $w_i$ take arbitrary real values.
 
 To solve MaxCut via VQE, a graph $G$ is encoded in the Ising model Hamiltonian

\begin{equation}
    H = \sum_i \omega_i \sigma_{ia} \sigma_{ib},
    \label{eq:H}
\end{equation}

\noindent where $\omega_i$ remains unchanged from the MaxCut objective function and $v_{ia}, v_{ib} \rightarrow \sigma_{ia}, \sigma_{ib}$ for Pauli-Z spin operators $\sigma_{ia}, \sigma_{ib}$. Maximizing the cut of $G$ is then equivalent to minimizing the loss function 

\begin{align}
    & \Lambda_t = \Lambda(\hat{\theta}_t) = \langle \mathbf{0}| (U_t^\dagger | H |U_t)| \mathbf{0} \rangle \\
    & = \sum_i \omega_i \langle \sigma_{ia} \sigma_{ib} \rangle_t = \sum_i \mu^i_t \nonumber,
\end{align}

\noindent where $\mu^i_t$ are the expectation values of the quadratic MaxCut terms. VQE circuit training updates parameters $\hat{\theta}$ via gradient descent on $\Lambda_t$ (Fig.\ \ref{fig:1}), where the gradient of any $\theta_t^k \in \hat{\theta}_t$ can be calculated as $\nabla_k \Lambda(\hat{\theta}_t) = \left( \Lambda(\hat{\theta}_t + \epsilon \hat{k}) - \Lambda(\hat{\theta}_t - \epsilon \hat{k}) \right) / 2 \epsilon $ by finite difference. As $\nabla \Lambda(\hat{\theta}_t) \rightarrow 0$ in the vicinity of both global \textit{and local} minima, VQE training is prone to stagnation at suboptimal solutions.

\section{Results}
\label{sec:results}

In this section, we present our novel method for enhancing the performance of VQAs with classical MCMCs, a technique that we dub MCMC-VQA. We start by briefly reviewing traditional MCMC, focusing on the Metropolis-Hastings algorithm. Then, we introduce MCMC-VQA, derive its behavior, and verify our findings with numerical simulations.

\subsection{MCMC-VQA Method}
\label{subsec:MCMC-VQ}

\begin{figure*}
\includegraphics[width=0.9\textwidth]{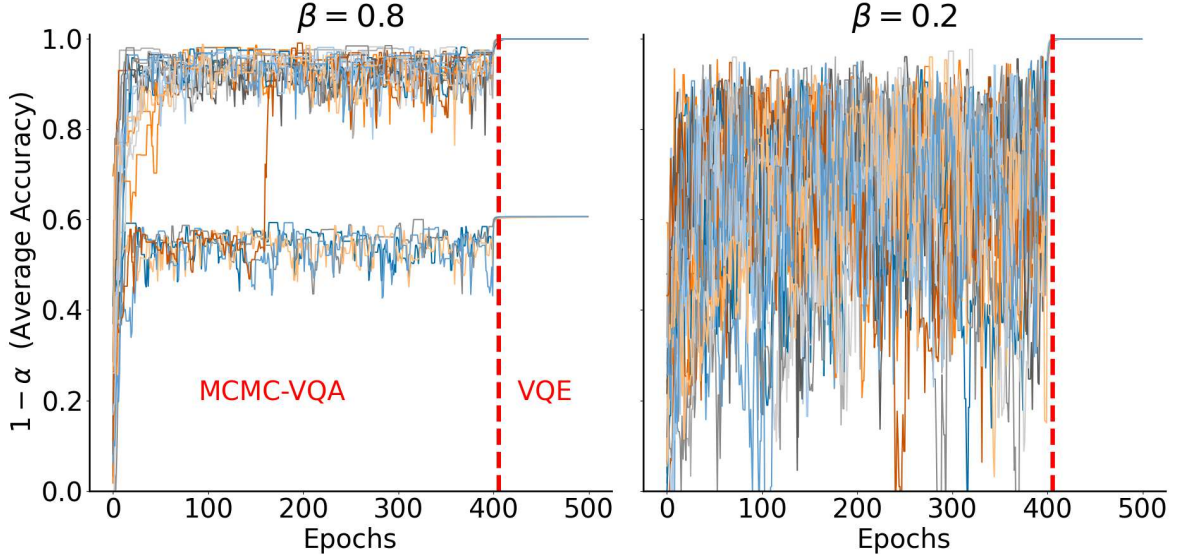}
\caption{Example trajectories with inverse thermodynamic temperature $\beta=0.8$ (left) and $\beta=0.2$ (right). Four-hundred MCMC-VQA epochs (Markovian epochs) are followed by a closing sequence of VQE epochs (beginning at red dashed line), which is initialized with the best parameters $\hat{\theta}_\text{min}$ found during the Markov process. At lower temperature ($\beta=0.8$), trajectories become trapped in local minima and reaching ergodicity is a lengthy process. Conversely, the high-temperature ($\beta=0.2$) trajectories rapidly reach burn-in, generating $\hat{\theta}_\text{min}$ that lead to near perfect convergence during the VQE closing sequence. See Sec.\ \ref{sec:methods} for simulation details.}
\label{fig:2}
\end{figure*}

 MCMC algorithms, such as Metropolis-Hastings, combine the randomized sampling of Monte-Carlo methods with the Markovian dynamics of a Markov chain in order to randomly sample from a distribution that is difficult to characterize deterministically \cite{Metropolis1953}. MCMC is particularly useful for approximations in high-dimensional spaces, where the so-called ``curse of dimensionality'' can make techniques such as random sampling prohibitively slow \cite{Geyer1992}. The core merit of MCMC techniques is their ergodicity, which guarantees that all states of the distribution are eventually sampled in a statistically representative way, regardless of which initial point is chosen. This representative sample is known as the unique stationary distribution $\pi$. In particular, any Markov chain that is both irreducible (each state has a non-zero probability of transitioning to any other state) and aperiodic (not partitioned into sets that undergo periodic transitions) will provably converge to its unique stationary distribution $\pi$, from which it samples ergodically \cite{Brooks1998}. The mathematical properties of ergodic Markov chains are well-studied, including analytic bounds for solution quality and mixing time (number of epochs) \cite{Montenegro2006, McNew2011}.

In order to obtain $\pi$ for a distribution of interest, Metropolis-Hastings specifies the transition kernel $P(x'|x)$, which is the probability that state $x$ transitions to state $x'$. Typically, the Markov process is defined such that transitions satisfy the detailed balance condition:

\begin{equation}
P(x) P(x'|x) = P(x') P(x|x').
\label{eq.detailed_balance}
\end{equation}

\noindent When Eq.\ \ref{eq.detailed_balance} holds, the chain is said to be reversible and is guaranteed to converge to a stationary distribution. $P(x'|x)$ can be factored into two quantities

\begin{equation}
    P(x'|x) = G(x'|x) A(x'|x),
\end{equation}

\noindent where $G(x'|x)$ is the proposal distribution, or the conditional probability of proposing state $x'$ given state $x$, and $A(x'|x)$ is the acceptance distribution, or the probability of accepting the new state $x'$ given state $x$. To satisfy Eq.\ \ref{eq.detailed_balance}, the acceptance distribution is defined as

\begin{equation}
    A(x'|x) = \min \left(1, \frac{P(x')G(x|x')}{P(x)G(x'|x)} \right).
\end{equation}

\noindent Note that as only the ratio $P(x') / P(x)$ is considered, the probability distribution need not be normalized. To determine whether the candidate state $x'$ or the current state $x_t$ should be used as the future state $x_{t+1}$, a sample $u$ is drawn from the uniform distribution $U(0,1)$. If $A(x'|x_t) \geq u$, then $x_{t+1} = x'$ and we say that the candidate state $x'$ is accepted. Otherwise, $x_{t+1}=x_t$ and we say that $x'$ is rejected.

We now present the MCMC-VQA method. Fig.\ \ref{fig:1} contains a diagram of the algorithm (blue). In particular, we focus on an ergodic Metropolis-Hastings algorithm, which is guaranteed to sample states near global minima. We outline the algorithm both idealistically and experimentally, prove its ergodicity and convergence, and verify these findings with numerical simulations.

As we seek the lowest energy eigenstate when solving MaxCut via VQE, we define $P(\hat{\theta})$ as the Boltzmann distribution

\begin{equation}
P(\hat{\theta}_a) = \exp \left( -\beta \Lambda_a \right) / Z, \hspace{0.8cm} Z = \sum_i \exp \left( -\beta \Lambda_i \right),
\label{eq:P}
\end{equation}

\noindent such that a state's probability increases exponentially with decreasing loss function.

To calculate the proposal distribution $G(\hat{\theta}'|\hat{\theta}_t)$, we must consider the sampling statistics of VQAs. Due to quantum uncertainty, a measurement $m_i^r(\hat{\theta}_t)$ of operators $\omega_i \sigma_{ia} \sigma_{ib}$ from Eq.\ \ref{eq:H} is a sample from a distribution with mean $\mu^i_t$ and variance

\begin{equation}
    (\Delta^i_t)^2 = \omega_i^2 [\langle (\sigma_{ia} \sigma_{ib})^2 \rangle_t - \langle \sigma_{ia} \sigma_{ib} \rangle_t^2 ] = \omega_i^2 [1 - (\mu^i_t)^2].
\end{equation}

\noindent The Central Limit Theorem asserts that, assuming at least $M \gtrsim 30$ independent and identically distributed measurements $m_i^r(\hat{\theta}_t)$, an estimate of the loss function $\Lambda_t$ is the statistic $l_t \sim \mathcal{N}\left(\Lambda_t, \hspace{0.05cm} (\Delta^\Lambda_t)^2 \right)$, where $(\Delta^\Lambda_t)^2 = \sum_i (\Delta^i_t)^2 / M$ \cite{kim2015t, Kwak2017}. Similarly, $\forall \theta_t^k \in \hat{\theta}_t$ and assuming small parameter shifts $\epsilon$, the gradient $\nabla_k \Lambda_t = \left( \Lambda(\hat{\theta}_t + \epsilon \hat{k}) - \Lambda(\hat{\theta}_t - \epsilon \hat{k}) \right) / 2 \epsilon $ is the statistic 
$d_k l_t \sim \mathcal{N}\left( \nabla_k \Lambda_t , \hspace{0.2cm} [ \Delta_{\Lambda}^2(\hat{\theta}_t + \epsilon \hat{k}) + \Delta_{\Lambda}^2(\hat{\theta}_t - \epsilon \hat{k}) ] / 4 \epsilon^2 \right)$. The variance of this distribution can be simplified by noting that to first order in $\epsilon$, the parameter shifted Pauli operators are $\sigma_{ia}^{\pm k} = \sigma_{ia}(\hat{\theta}\pm \epsilon \hat{k}) = \sigma_{ia} \pm \iota_{iak}$, where $\sigma_{ia} = \sigma_{ia}(\hat{\theta})$ and $\iota_{iak} = (\partial \sigma_{ia} / \partial \theta^k) \epsilon$. We can then simplify the sum $\Delta_i(\hat{\theta}_t + \epsilon \hat{k})^2 + \Delta_i(\hat{\theta}_t - \epsilon \hat{k})^2 = 2 \Delta_i(\hat{\theta}_t)^2$ by noting that

\begin{subequations}
 \begin{align}
  & \Delta_i(\hat{\theta}_t + \epsilon \hat{k})^2 = \langle (\omega_i \sigma_{ia}^{\pm k} \sigma_{ib}^{\pm k})^2 \rangle - \langle \omega_i \sigma_{ia}^{\pm k} \sigma_{ib}^{\pm k} \rangle^2, \\
  & \langle (\sigma_{ia}^{+ k} \sigma_{ib}^{+ k})^2 \rangle + \langle (\sigma_{ia}^{- k} \sigma_{ib}^{- k})^2 \rangle = 2 + \mathcal{O}(\iota^2), \label{eq1} \\
  & \langle \sigma_{ia}^{+k} \sigma_{ib}^{+k} \rangle^2 + \langle \sigma_{ia}^{-k} \sigma_{ib}^{-k} \rangle^2 = 2 \langle \sigma_{ia} \sigma_{ib} \rangle + \mathcal{O}(\iota^2). \label{eq2}
 \end{align}
\end{subequations}

\noindent Now, up to first order in $\iota$, we can derive the gradient's distribution

\begin{equation}
d_k l_t \sim \mathcal{N}\left( \nabla_k \Lambda_t , \hspace{0.1cm} \Delta_{\Lambda}^2(\hat{\theta}_t) / 2 \epsilon^2 \right).
\end{equation}

Standard gradient descent would propose the candidate state $\hat{\theta}' = \hat{\theta} - \eta \nabla \Lambda_t$, however MCMC-VQA adds a normally distributed random noise term $\Theta_r \sim \mathcal{N}(0,1)$ with scale parameter $\xi$ in order to expand the support of the proposal distribution $G(\hat{\theta}'|\hat{\theta}_t)$. This specifies

\begin{widetext}
\begin{equation}
    G(\hat{\theta}'|\hat{\theta}_t) = \prod_k G(\hat{\theta}'|\hat{\theta}_t)_k, \hspace{0.4cm} G(\hat{\theta}'|\hat{\theta}_t)_k = \text{pdf}\left[\mathcal{N} \left(\eta \nabla_k \Lambda (\hat{\theta}_t), \hspace{0.2cm} \xi^2 + \eta^2 \frac{(\Delta^\Lambda_t)^2}{2 \epsilon^2} \right) \right] \left( \hat{\theta}_t - \hat{\theta}' \right),
    \label{eq:proposal_distribution}
\end{equation}
\end{widetext}

\noindent where the notation $\text{pdf}\left[\mathcal{N}\left(\mu, \sigma^2 \right) \right] (x)$ denotes the probability density function at point $x$ of a normal distribution with mean $\mu$ and variance $\sigma^2$. It follows that the acceptance distribution is given by

\begin{equation}
    A(\hat{\theta}'|\hat{\theta}_t) = \min \left(1, \frac{P(\hat{\theta}')G(\hat{\theta}_t|\hat{\theta}')}{P(\hat{\theta}_t)G(\hat{\theta}'|\hat{\theta}_t)} \right).
\end{equation}

\noindent We note that $G(\hat{\theta}_t|\hat{\theta}')$ is obtained by simply exchanging $\hat{\theta}_t$ and $\hat{\theta}'$ in Eq.\ \ref{eq:proposal_distribution}. A random uniform sample $u \sim U(0,1)$ is then drawn for comparison, such that $\hat{\theta}_{t+1} = \hat{\theta}'$ if $A(\hat{\theta}'|\hat{\theta}_t) > u$ and $\hat{\theta}_{t+1} = \hat{\theta}_t$ otherwise.

After $T_\text{MC}$ epochs of the above Markovian process, MCMC-VQA implements a short series of traditional VQA epochs for rapid convergence to the nearest minimum. In particular, these closing VQA epochs are initialized with $\hat{\theta}_\text{min}$, the parameter set of lowest eigenvalue $\Lambda_\text{min}$ found during the Metropolis-Hastings phase. In this manner, MCMC-VQA can be considered a ``warm starting'' procedure \cite{beaulieu2021max, egger2021warm, van2021}, but with ergodic guarantees.

Example MCMC-VQA trajectories are shown in Fig.\ \ref{fig:2} with inverse thermodynamic temperatures $\beta=0.8$ and $\beta=0.2$. The details of all simulations are given in Sec.\ \ref{sec:methods}. Our algorithm combines the gradient descent-based optimization of VQE with a Markovian process that escapes local minima. Such exploration is significantly greater at the higher-temperature $\beta=0.2$, where rather than settling into distinct loss function basins from which escape is relatively rare, the trajectories display the trademark ``burn-in'' behavior of ergodic Markov chains. By the time that the closing VQE epochs are applied, the ergodic $\beta=0.2$ MCMC-VQA chains have sampled states sufficiently near the global minimum and converge to the groundtruth nearly uniformly.

\begin{figure*}
\includegraphics[width=0.9\textwidth]{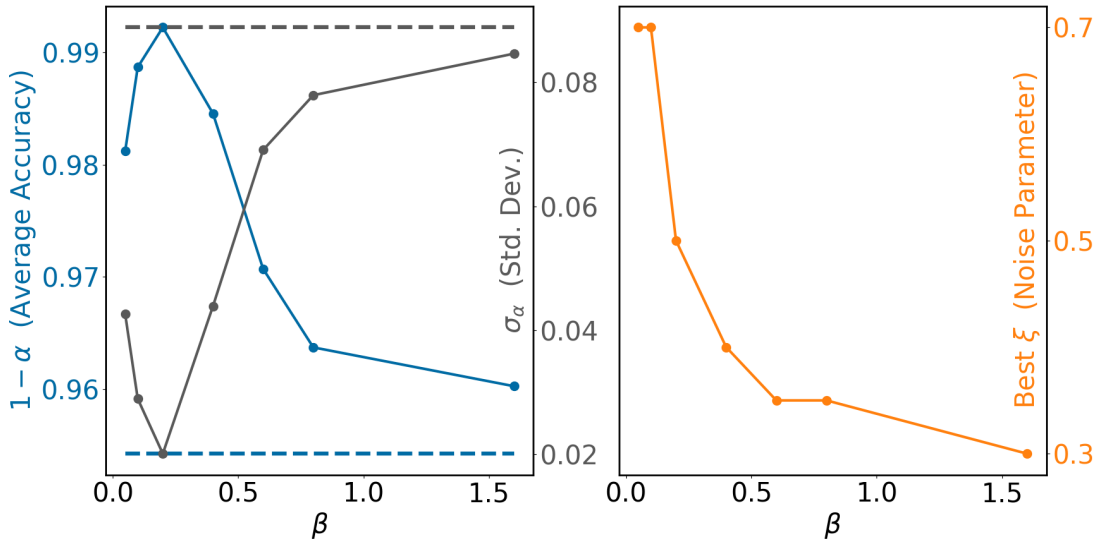}
\caption{(Left, blue) Average MCMC-VQA accuracy ($1-\alpha$, for average error $\alpha$) vs inverse thermodynamic temperature $\beta$. Nearly perfect average accuracy is obtained for properly tuned hyperparameter $\beta$ (here, $\beta \approx 0.2$). At low temperature (large $\beta$), the algorithm mixes slowly, only partially approximating ergodicity in $T_\text{MC}=400$ Markovian epochs. This partial convergence results in lower accuracy, which approaches that of traditional VQE (blue dashed line) in the limit of large $\beta$. Conversely, for high temperature (small $\beta$), the algorithm is insufficiently biased towards low-energy solutions, which renders its gradient descent inefficient and reduces its accuracy. (Left, gray) The standard deviation of MCMC-VQE accuracy vs $\beta$. Higher standard deviation directly corresponds with lower accuracy. As discussed above, at high $\beta$, this is due to runs trapped in local minima (see Fig.\ \ref{fig:2}), while at low $\beta$, this stems from the lack of energy-preferred convergence. (Left) Optimal value of $\xi$ vs $\beta$, where $\xi$ is the gradient descent noise parameter ($\hat{\theta}' = \hat{\theta} - \eta \nabla \Lambda_t + \xi \Theta_r$) and each trajectory undergoes $T_\text{MC}=400$ Markovian epochs. As larger temperatures generate more permissive acceptance distributions $A(\hat{\theta}'|\hat{\theta})$, higher $\xi$
values lead to more efficient mixing in the low-$\beta$ limit. See Sec.\ \ref{sec:methods} for simulation details.}
\label{fig:3}
\end{figure*}

Fig.\ \ref{fig:3} (left) displays the average accuracy $1-\alpha$ (where $\alpha$ is the average error, blue), and standard deviation (gray) of MaxCut solutions with MCMC-VQA as a function of $\beta$. Dashed lines represent the performance of traditional VQE on the same set of graphs and circuit ansatz. We note that all simulated $\beta$ values outperform traditional VQE. Until $\beta \sim 0.2$, higher temperature MCMC-VQA chains have higher accuracy and better convergence, as their more permissive temperature parameter biases the acceptance distribution towards accepting the candidate states. However, performance decreases at very high temperatures, for which the MCMC-VQA chains are no longer appreciably biased towards energy minimization and the algorithm becomes more like random sampling than intrepid gradient descent. Likewise, the optimal amount of parameter update noise $\xi$ is inversely proportional to $\beta$ (Fig.\ \ref{fig:3}, right), as higher temperatures permit more radical deviations from standard gradient descent.

\subsection{Implementation of MCMC-VQA on Quantum Hardware}
\label{subsec:experimental_considerations}

As discussed above, the loss function $\Lambda_t$ is not precisely determined on actual quantum hardware, but rather estimated as a statistic $l_t = \sum_i q^i_t$, where $q^i_t = \frac{1}{M} \sum_{r=1}^{M} m_i^r(\hat{\theta}_t)$. As a result, the variance of a single observable measurement $(\Delta^i_t)^2$ is estimated by $(\delta^i_t)^2 = \omega_i^2 [1 - (q^i_t)^2]$, while that of the total loss function $(\Delta^\Lambda_t)^2$ is estimated by $(\delta^\Lambda_t)^2 = \sum_i (\delta^i_t)^2 / M = \sum_i \omega_i^2 [1 - (q^i_t)^2] / M$, for $M$-measurements per observable. Alternatively, the variances could be directly estimated from the standard deviations of expectation value statistics. We then define $a(\hat{\theta}'|\hat{\theta}_t)$, the acceptance distribution on quantum hardware, as

\begin{subequations}
 \begin{align}
  & a(\hat{\theta}'|\hat{\theta}_t) = \min \left(1, \frac{p(\hat{\theta}')g(\hat{\theta}_t|\hat{\theta}')}{p(\hat{\theta}_t)g(\hat{\theta}'|\hat{\theta}_t)} \right), \\
  & p(\hat{\theta}) \propto \exp(- \beta l_t), \\
  & g(\hat{\theta}'|\hat{\theta}_t) = \prod_k g(\hat{\theta}'|\hat{\theta}_t)_k, \\
  & g(\hat{\theta}'|\hat{\theta}_t)_k = \text{pdf}\left[\mathcal{N} \left(\eta d_k l_t, \hspace{0.2cm} \xi^2 + \eta^2 \frac{(\delta^\Lambda_t)^2}{2 \epsilon^2} \right) \right] \left( \hat{\theta}_t - \hat{\theta}' \right).
 \end{align}
\end{subequations}

\noindent MCMC-VQA does not increase the quantum complexity of VQAs (number of operations carried out on quantum hardware), as the measurements to estimate $\Lambda(\hat{\theta})$ are carried out in the typical way. Moreover, the acceptance distribution and its components are computed classically with simple arithmetic.

\begin{figure*}
\includegraphics[width=0.9\textwidth]{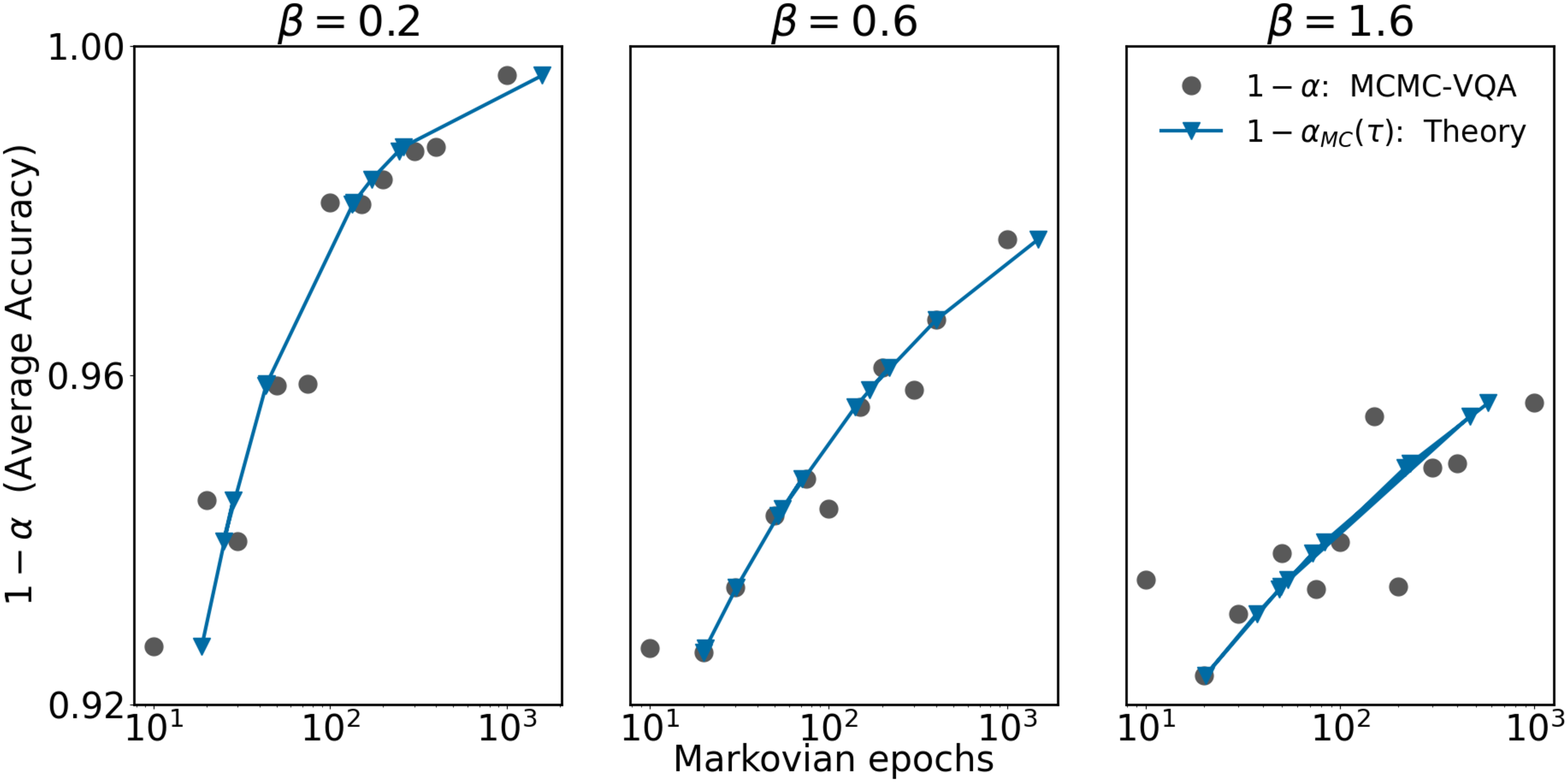}
\caption{Average accuracy vs Markovian epochs for three different $\beta$ values. Gray dots are the average MCMC-VQA accuracy $1-\alpha$, and blue curves are a least squares fit of this data to the analytical accuracy of an ergodic Markov chain $1 - \alpha_\text{MC}(\tau)$, with theoretical mixing time $\tau$ (see Eq.\ \ref{eq:mixing_time}). The analytical time-dependence of $\alpha_\text{MC}$ matches the observed scaling of $\alpha$, affirming that MCMC-VQA is an ergodic Markov chain, and thus guaranteeing convergence to the global minimum. Furthermore, the ratio of observed scale parameters between MCMC-VQA simulations with different $\beta$ values is consistent with the analytic dependence $\tau \propto \ln(1/\sqrt{\pi^*})$ (Eq.\ \ref{eq:mixing_time}) on the least likely state $\pi^* \propto \exp(-\beta \Lambda_\text{max})$ (Eq.\ \ref{eq:P}). This functional dependence on temperature further supports our claims of ergodically sampling from $P(\hat{\theta})$ and thus deterministically converging to the global minimum.}
\label{fig:4}
\end{figure*}

\subsection{Proof of Ergodicity}
\label{subsec:ergodicity}
If a Metropolis-Hastings algorithm is \textit{irreducible} and \textit{aperiodic}, then the resulting Markov chain is provably ergodic \cite{Brooks1998}. That is, it will explore all areas of the probability distribution, converging on average to the Markov process' unique stationary distribution, which includes the global minimum of the solution space. Moreover, as we have chosen to sample from the Boltzmann distribution of the loss function, we sample from states near optimal solutions with exponentially higher probability.

\subsubsection{Irreducibility}

The VQA Metropolis-Hastings Markov chain is irreducible if $\forall \hat{\theta}_a, \hat{\theta}_b, \hspace{0.2cm} \exists T, \{\hat{\theta}_1, \hat{\theta}_2, ....., \hat{\theta_T} \}$ such that

\begin{equation}
    p(\hat{\theta}_1|\hat{\theta}_a) p(\hat{\theta}_b|\hat{\theta}_T) \prod_{i=1}^{T-1} p(\hat{\theta}_{i+1}|\hat{\theta}_i) > 0.
    \label{eq:irreducible}
\end{equation}

\noindent That is, the Markov chain is irreducible if, for any two points in parameter space $\hat{\theta}_a, \hat{\theta}_b$, there exists a series of transitions of any length $T$ such that $\hat{\theta}_a \rightarrow \hat{\theta}_b$ with non-zero probability \cite{Daskalakis}. While this definition of irreducibility is sufficient, we will instead focus on the yet more powerful condition of \textit{strong} irreducibility. A Markov chain is strongly irreducible iff

\begin{equation}
g(\hat{\theta}_a|\hat{\theta}_b) > 0, \forall \hat{\theta}_a, \hat{\theta}_b,
\label{eq:strongly_irreducible}
\end{equation}

\noindent meaning that all points in parameter space have a non-zero probability of transitioning to all other points \cite{Whiteley}. This condition is then equivalent to

\begin{widetext}
\begin{equation} 
    g(\hat{\theta}_b|\hat{\theta}_a)_k = \frac{(2 \pi)^{-1/2}}{\sqrt{\xi^2 + \eta^2 (\delta^\Lambda_a)^2 / 2 \epsilon^2}} \exp\left[ \frac{ -\left( \theta_a^k - \theta_b^k - \eta d_k l_a \right)^2}{2 \left( \xi^2 + \eta^2 (\delta^\Lambda_a)^2 / 2 \epsilon^2 \right)} \right] > 0, \hspace{0.2cm} \forall k,
\end{equation}
\end{widetext}

\noindent where we note that $\delta_\Lambda^2(\hat{\theta}_t) \propto 1/M$.

Eq.\ \ref{eq:strongly_irreducible} is satisfied, at least technically to some tolerance, $\forall \hat{\theta}_a, \hat{\theta}_b$. Although $g(\hat{\theta}_b|\hat{\theta}_a)_k$ may become very small, it will generally retain a non-zero probability for virtually all transitions, and the chain will be strongly irreducible, albeit perhaps slow to convergence. More precise arguments can be made in the limit of large $\xi$, where to first order in small $1/\xi$,  $g(\hat{\theta}_b|\hat{\theta}_a)_k \rightarrow 1/\sqrt{2 \pi} \xi$ and all transitions become equally likely. While this extreme $\xi$ limit is too random to result in efficient gradient descent, it illustrates a concrete transition to irreducibility with increasing $\xi$. Moreover, due to the uncertainty introduced by finite statistics $d_k l_a$ and $(\delta^\Lambda_a)^2$, sampling of the proposition kernel $g(\hat{\theta}_b|\hat{\theta}_a)_k$ can allow for otherwise unlikely transitions. 

\subsubsection{Aperiodicity}

In the case of strong irreducibility argued above (Eq.\ \ref{eq:strongly_irreducible}), aperiodicity is automatically satisfied. Assuming only the weaker irreducibility of Eq.\ \ref{eq:irreducible}, it is sufficient to show that \cite{Daskalakis}

\begin{widetext}
\begin{equation}
    a(\hat{\theta}_a|\hat{\theta}_a) g(\hat{\theta}_a|\hat{\theta}_a) = g(\hat{\theta}_a|\hat{\theta}_a) = \frac{(2 \pi)^{-1/2}}{\sqrt{\xi^2 + \eta^2 (\delta^\Lambda_a)^2 / 2 \epsilon^2}} \exp\left[ \frac{ -\left(\eta d_k l_a \right)^2}{2 \left( \xi^2 + \eta^2 (\delta^\Lambda_a)^2 / 2 \epsilon^2 \right)} \right] > 0.
    \label{eq:aperiodicity}
\end{equation}
\end{widetext}

\noindent As long as $\eta \not\gg \xi$, Eq.\ \ref{eq:aperiodicity} holds for all but singular points $\hat{\theta}_a$.

\subsection{Mixing Time}

The mixing time $\tau$ of a Markov chain is the number of epochs required to reach a certain threshold of convergence. For an ergodic, discrete-time Markov chain, $\tau$ is analytically bounded by

\begin{equation}
    \tau \leq \frac{2}{\Phi^2} \ln\left(\frac{1}{\alpha_\text{MC} \sqrt{\pi^*}} \right),
    \label{eq:mixing_time}
\end{equation}

\noindent where $\alpha_{MC} = |S-\pi|$ is the distance between the Markov chain's sampled distribution $S$ and the true stationary distribution $\pi$, $\pi^*$ is the probability of the least likely (maximum energy) state of $\pi$, and $\Phi$ is the conductance or ``Cheeger constant'' of the Markov process \cite{Montenegro2006}. The conductance can be understood as the minimum of normalized ergodic flows between all possible partitions of the state space.

Fig.\ \ref{fig:4} demonstrates that the performance of MCMC-VQA is consistent with the theoretical predictions of ergodic Markov chains (Eq.\ \ref{eq:mixing_time}). That is, the time dependence of MCMC-VQA optimization error $\alpha$ follows the same $\ln(1 / \alpha)$ scaling as the distribution distance $\alpha_\text{MC}$ in Eq.\ \ref{eq:mixing_time}. Moreover, least-squares analysis of Fig.\ \ref{fig:4} data reveals a $\beta$-dependent scale factor that is proportional to $\ln(1 / \sqrt{\pi^*})$, which is consistent with the Boltzmann distribution $p(\hat{\theta}_a) \propto \exp(-\beta \Lambda_a)$ from which our method samples. This temperature-dependent time-complexity further verifies that MCMC-VQA is an ergodic Markov process that successfully samples from the target distribution.

\section{Numerical Simulations}
\label{sec:methods}

The simulations in this work are done using a modified version of TensorLy-Quantum, an open-source software package for quantum circuit simulation using factorized tensors \cite{Patti2021_tlq,TLQ}. TensorLy-Quantum specializes in exact tensor contraction, such that the simulations are carried out without truncation or approximation.

The MaxCut instances optimized in this work are generated from ten graphs. Each graph has ten vertices and an equal number of randomly selected edges, which are randomly generated from the unit normal distribution. Such graphs are equivalent to the Gilbert model of random graphs \cite{gilbert1959random}. The number of edges was chosen to be equal to that of vertices as this ratio is observed to pose high difficulty for random MaxCut problems of this model \cite{coppersmith2004random, luczak1990equivalence}.
 
 All numerical simulations in this work are done using the graphs described above, with twenty randomly initialized runs completed for each graph. The quantum circuits use one parameterized rotation per vertex.  We illustrate our work using circuits with relatively few parameters, because their optimization landscape is especially nonconvex and thus prone to convergence in local minima \cite{Lee2021}, however MCMC-VQA can be used with arbitrary parameterization. The circuit gates are alternated between a layer of single-qubit parameterized rotations (angles $\hat{\theta}$) about the $y$-axis and a layer of two-qubit control-Z gates. For each method (VQE or MCMC-VQA) and set of hyperparameters, a variety of learning rates are scanned so that numerical comparisons could be drawn against the optimal performance of each algorithm. All VQE sequences consisted of 100 epochs. Fig.\ \ref{fig:2} shows an ensemble of trajectories whereas Figs.\ \ref{fig:3} and \ref{fig:4} is the average over the optimal learning rate for all ten graphs and 20 random initializations. For simplicity, we take the large $M$ limit, assuming many measurements and precise expectation values.

\section{Discussion}
\label{sec:disc}

In this work, we have introduced MCMC-VQA: a novel variational quantum algorithm that harnesses classical Makov chains to obtain analytic convergence gaurantees for parameterized quantum circuits. As ergodic Markov chains representatively sample a target probability distribution, they identify regions near the global minimum with high probability. We present MCMC-VQA, both from a theoretical and practical perspective, prove its ergodicity, and derive its time-complexity (mixing time) as a function of both accuracy and inverse thermodynamic temperature. Focusing on MaxCut optimization within the VQE framework due to its plentiful local minima and employing a reversible Metropolis-Hastings Markov process, we demonstrate the ergodicity of our method, the validity of our analytical findings, and the capacity of MCMC-VQA to not only outperform traditional VQAs, but to do so with up to perfect and deterministic convergence.

In future research, MCMC-VQA should be studied for a variety of different applications, quantum algorithms, and Markov processes. In addition to quantum optimization, VQAs have been employed to address a myriad of topics in both quantum chemistry \cite{mcardle2018quantum, kandala2017hardware, Grimsley2019} and condensed matter physics \cite{ritter2019near, vogt2020preparing, Zhang2021}. Moreover, even simple quantum Hamiltonians, such as the transverse field Ising model, are known to acutely struggle with premature convergence to local, rather than global, minima. Similarly, our technique could be extended to QAOA \cite{farhi2014quantum} or any of the numerous VQAs that have been proposed in recent years. Finally, tens of MCMCs have been devised over the past 70 years, each with their own advantages, with variations featuring Gibbs sampling \cite{gelfand2000gibbs}, parallel tempering \cite{earl2005parallel}, and independence sampling \cite{Hastings1970}. These methods could be substituted for Metropolis-Hastings in order to produce algorithms with lower computational overhead and faster mixing times. In short, varieties of MCMC-VQA can be developed for a broad spectrum of variational quantum algorithms to both improve and guarantee performance.

\section*{Acknowledgements}
O.S. likes to thank Katie Pizzolato for accommodating HPC resource requests on IBM Cloud. This work was done during T.L.P.'s internship at IBM Quantum, for which T.L.P. thanks Katie Pizzolato and the entire IBM Quantum team. S.F.Y. would like to acknowledge funding by NSF and AFOSR.

\bibliography{references}

\appendix

\end{document}